\documentclass[12pt,preprint]{aastex}

\newcommand{\etal}{{\it et al.}}
\slugcomment{Accepted for Publication in ApJL}
\shorttitle{Low frequency coherence break}
\shortauthors{Jian Feng Ji \etal}

\begin{document}

\title{Low Frequency Coherence Break in The Soft X-ray State of GRS 1915+105}
\author{Jian Feng Ji\altaffilmark{1}, Shuang Nan Zhang\altaffilmark{2,3,4,5},Jin Lu Qu\altaffilmark{5} and Ti Pei Li\altaffilmark{2,5}}

\altaffiltext{1}{Department of Engineering Physics, Tsinghua University, Beijing 100084, P.R.China; jjf00@mails.tsinghua.edu.cn}
\altaffiltext{2}{Department of Physics \& Center for Astrophysics, Tsinghua University, Beijing 100084, P.R.China; zhangsn@mail.tsinghua.edu.cn, litp@mail.tsinghua.edu.cn}
\altaffiltext{3}{Physics Department, University of Alabama in Huntsville, Huntsville, AL 35899; zhangsn@email.uah.edu}
\altaffiltext{4}{Space Science Laboratory, NASA Marshall Space Flight Center, SD50, Huntsville, AL 35812, USA}
\altaffiltext{5}{Laboratory for Particle Astrophysics, Institute of High Energy Physics, The Chinese Academy of Sciences, Beijing 100039, P.R. China; qujl@mail.ihep.ac.cn}

\begin{abstract}
We present results from the analysis of X-ray power density spectra and coherence when
 GRS 1915+105 is in soft states.
We use three data sets that belong to $\mu$, $\phi$ and $\delta$ classes in Belloni \etal  (2000).
We find that the power density spectra appear to be complex, with several features between 0.01 and 10 Hz. The coherence deviates from unity above a characteristic frequency.
We discuss our results in different models.
The corona size in the sphere-disk model implied by this break frequency is on the order of  10$^4$ GM/c$^2$,
which is unphysical. Our results are more consistent with the prediction of the model of a
planar corona sustained by magnetic flares, in which the characteristic frequency is
associated with the longest time-scale of an individual flare, which is about eight seconds.
\end{abstract}


\section{Introduction}

The X-ray source GRS~1915+105 was discovered by the GRANAT observatory in 1992
(Castro-Tirado \etal  1992). It is one of the several galactic
objects observed to produce superluminal jets (Mirabel \etal  1994).
As the first galactic superluminal jet source, the so-called ``microquasar", GRS~1915+105 is a unique
and very important astrophysical laboratory for relativistic astrophysics (Mirabel \etal  1998, 1999).
Zhang \etal  (1997) first suggested that the black hole in this system is highly spinning; subsequent studies
on the high frequency QPO phenomena from this source tend to support the spinning black hole model for this source
(see, e.g., Cui \etal  1998, Remillard \etal  2002), though the case is not settled yet.

The spectral type of the companion has been recently identified as a K--M III star
(Greiner \etal  2001), thus the source is classified as a low mass X-ray binary. The estimated mass of the black hole
($\sim 14M_{\odot}$) in this system is significantly higher than the black hole masses in most of other stellar mass black hole systems (Greiner \etal  2001).
This system thus may have a unique evolutionary history and this also poses a major challenge to the theories of massive
star evolution.

GRS~1915+105 displays rich QPO phenomena.
The 328 Hz QPO (Remillard \etal  2002), the highest QPO in this source,
is believed to be associated with the innermost stable radius of accretion disk.
A stable $\sim$ 67 Hz QPO (Morgan \etal  1997) accompanied with a 40 Hz QPO (Strohmayer 2001) has also been observed.
The 0.5-10 Hz QPOs, observed only during the hard state,
are probably linked to the properties of the accretion disk,
since their centroid frequency and fractional rms have been
reported to correlate with the thermal flux (Markwardt \etal  1999,
Trudolyubov \etal  1999, Reig \etal  2000) and apparent temperature (Muno \etal  1999).
At much lower frequencies (0.001-0.1 Hz), the source occasionally shows high-amplitude
QPOs or brightness ``sputters" (Morgan \etal  1997). These variations correspond probably to the disk instability.

The 67 Hz QPO shows a marked hard lag when the low frequency (well below 1 Hz) QPO shows a complex hard/soft lag
structure ( Cui \etal  1999). The 0.5-10 Hz QPO shows a complex lag behavior (Reig \etal  2000). The lag between hard and soft photons
decreases as the frequency of the QPO increases, changing the sign of the lag for $\nu_{QPO}\geq 2$ Hz. Negative lags occur when the
power-law spectrum is soft and positive lags occurs when the spectrum is hard. The 0.5-10 Hz
QPO disappears when the PCA intensity is high ($\geq2000$ counts per second) (Reig \etal  2000).

Based on the X-ray color-color diagram, Belloni \etal  (2000) found that the complex X-ray variability of GRS~1915+105 can be reduced to transitions between three basic
spectral states, which they called ``A", ``B" and ``C". The spectrally soft states ``A" and ``B" correspond to an observable
inner accretion disk with different temperatures: in state ``B" the inner disk
temperature is higher compared to state ``A". For
the spectrally hard state ``C", the inner part of the accretion disk is either missing or just unobservable.

The coherence function between the light curves in two different energy bands measures how the photons in the
two energy bands are related (Vaughan \etal  1997). For black hole binaries, the coherence function
often appears to be around unity
over a wide frequency range (Vaughan \etal  1997; Cui \etal  1997; Nowak \etal  1999a; Nowak \etal  1999b), indicating
that high energy photons are closely related to low energy photons, or the low energy photons are the seed
photons of the high energy photons.
Reduced coherence was observed from Cyg X-1 when the source was in the transition state (Cui \etal  1997), indicating that the hard X-ray production
region was not stable during the transition state. Therefore the coherence function provides a
useful probe into the physical properties
of the hard X-ray production region.

The source's X-ray temporal properties change with the radio flux.
In the steady hard state, as the radio emission becomes brighter and more
optically thick, the rms of the 0.5-10 Hz QPO decreases,
the Fourier phase lags in the frequency range of 0.01-10 Hz change sign,
the coherence at low frequencies decreases, and the relative amount of low
frequency power in hard photons compared to soft photons decreases (Muno \etal  2001). Klein-Wolt \etal  (2002)
found a relation between the above mentioned state ``C" and the radio flux. During state ``C" a more or less continuous ejection
of relativistic particles takes place. The length of the state ``C" interval determines the strength of the radio emission. The separation between different episodes of state ``C" determined the shape of the radio light curve.

In this paper, we make use of the RXTE observations of GRS~1915+105 in soft states to
examine its timing properties further. We choose three data sets belonging to three
different classes according to Belloni \etal  (2000).
In section 3, we first present the Power Density Spectra (PDS) and
then we consider the coherence. Finally we summarize our results and discuss the implications of our results in two different models.

\section{Data And Analysis}

The data used in this work are retrieved from the public RXTE archive
(see table ~\ref{Tb.1}). They belong to the soft states without prominent 0.5-10 Hz QPOs,
belonging to the spectral classes $\phi$, $\mu$, and $\delta$ respectively (Belloni \etal  2000). Because within each classes
the temporal properties of the source may vary at different times, we first examined the data for each individual
observation. We found that the data for the first and third sets show consistent temporal properties within each set,
and therefore
in the following we combine the PDS and coherence of all observations within the first and third sets. For the second
set, we found that although the general shape of the PDS for all observations is the same, the fine features vary between
the three observations. We therefore present the results of each individual observation for the second set separately.

The Proportional Counter Array (PCA) data modes that we used are listed in Table ~\ref{Tb.2}.
``Good Time Intervals" (GTIs) were defined when the elevation angle was above $10^\circ$,
the offset pointing less than $0.02^\circ$, and the number of PCUs turning ``on" equals to 5.

The techniques that we used to calculate the coherence for the GRS 1915+105 are
discussed in Vaughan \etal  (1997) and Nowak \etal  (1999a). We applied eq.(8) of Vaughan \etal  (1997) to our data.
We use eq.(1) of Morgan \etal  (1997) throughout this work to estimate the deadtime-corrected Poisson noise level;
all averaged PDS we present have been subtracted by poisson noise level.

For the first and second data set, we performed FFT with two segment lengths, 1024 s and 256 s respectively.
Segments with data gaps of any duration were ignored. In these results, the frequency range of 0.00097-0.01 Hz is
computed from the 1024 s data segments, and the frequency range of 0.0039125-64 Hz is computed from the 256 s data segments. For the third data set, we only performed FFT with segments of 256 s because of the shorter exposure time. Throughout this work we have used a logarithmic frequency binning with a binning factor of 0.1.

We also divided our data into five energy bands. The ranges of these energy bands are listed in Table ~\ref{Tb.3} .

\section{Results}

\subsection{Power Density Spectra} \label{PDS}
\begin{figure}
\plotone{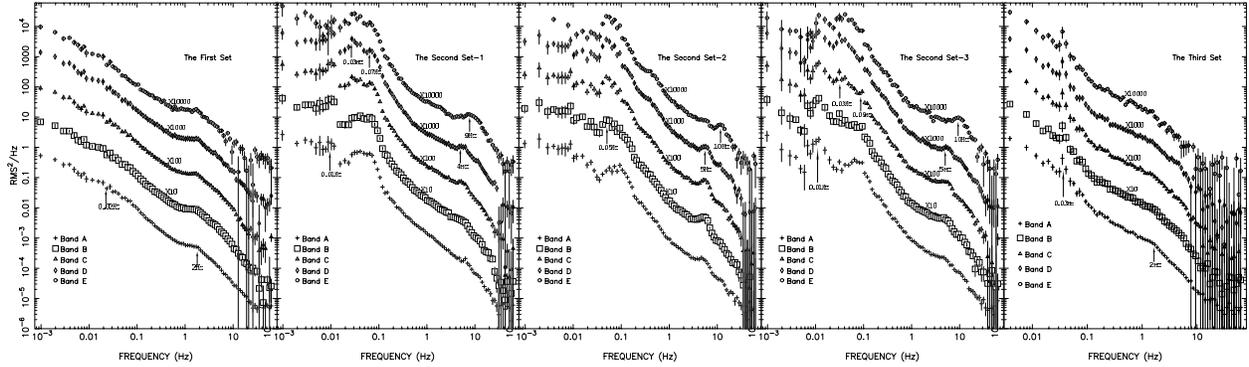}
\caption{PDS with associated uncertainties. All PDS are for the one-sided normalization of
Belloni \etal  (1990). \textit{Top left}: The first set. \textit{Top right}:
The second set. \textit{Bottom}: The third set. The PDS for bands ``B", ``C", ``D" and ``E" are multiplied by
factors of 10, 100, 1000 and 10000, respectively.\label{fig1}}
\end{figure}

In figure ~\ref{fig1}, we present the PDS of all three data sets in five energy bands. From the figure, we find:

1. The overall shape of the PDS for all three sets is power-law-like, superimposed with QPOs or broad features at
different frequencies, characteristic of black hole binaries in the soft state.

2. For the first data set, there is a knee at about 2 Hz. There may be a broad feature at about 0.02 Hz.

3. For the second data set, the PDS for each observation shown separately.
For the second set-1 and second set-3 , there are three QPOs at around 0.01 Hz, 0.03 Hz and 0.07 Hz.
For the second set-2 , there are two QPOs at around 0.01 Hz and 0.05 Hz. In all these observations,
there are two features at about 5 Hz and 10 Hz.
The 5 Hz feature increases with energy but decreases remarkably in the highest energy band.
However, the 10 Hz QPO is very weak in the low energy band but suddenly enhanced in the highest energy band.

4. For the third set, there is a QPO at about 0.03 Hz.

\subsection{Coherence} \label{COH}

\begin{figure}
\plotone{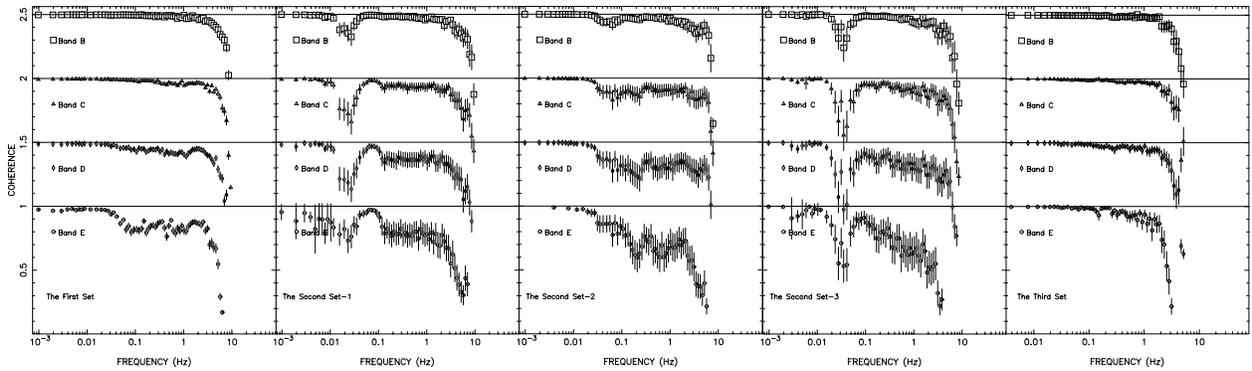}
\caption{Coherence functions of various energy bands. \textit{Top left}:
The first set. \textit{Top right}: The second set. \textit{Bottom}: The third set.
The coherence functions for bands ``D", ``C", ``B" are offset by 0.5, 1.0 and 1.5 respectively.\label{fig2}}
\end{figure}

In figure ~\ref{fig2}, we present the coherence for all energy bands for every data set.
All comparisons shown are relative to the band ``A". From the figure, we find:

1. For the first and second set the coherence between 0.03 Hz and 10 Hz becomes weak for higher energies.
For the third set, this trend is weaker.

2. For the first and second set, the coherence is remarkably close to unity below about 0.02 Hz for
all energy bands; whereas for the third set, the break frequency is about 1 Hz.

3. For the second set-1 and second set-3 there is a remarkable dip at about 0.3 Hz in the coherence curve;
the dip is deeper for higher energies.

4. For all data sets, the coherence deviates quickly from unity above about 10 Hz.

\section{Discussion and conclusions}

We have studies the temporal properties of the spectral classes $\phi$, $\mu$ and $\delta$ in the soft states of the microquasar
GRS~1915+105. We found in their PDS there are several low frequency QPOs or broad features between 0.01 to 10 Hz. For
the observations we have analyzed, the temporal properties for classes $\phi$ and $\delta$ do not show significant
variations across different observations. For the class $\mu$, we found significant variations on its temporal properties in
different observations.

The main results of this work are on the coherence function of different classes. For all three classes, the coherence function
shows a significant drop above 10 Hz; an energy dependent coherence break between 0.01 to 1 Hz also exists. In particular
for class $\mu$, there also exists a coherence dip at around 0.03 Hz, which corresponds to a dip between two broad
peaks on its power-density spectrum.

The quick loss of coherence above about 10 Hz seen for GRS~1915+105 in the soft state is similar to that seen
in Cyg X-1 and GX339-4 in the hard state; this may be due to some nonlinear processes at high frequencies (Nowak \etal  1999a).
However the coherence loss between 0.02 Hz and 10 Hz in GRS~1915+105 is different from that seen in Cyg X-1 and GX339-4.
This difference may be caused by the different spectral states when the sources were observed.

\begin{figure}[h]
\epsscale{0.3}
\plotone{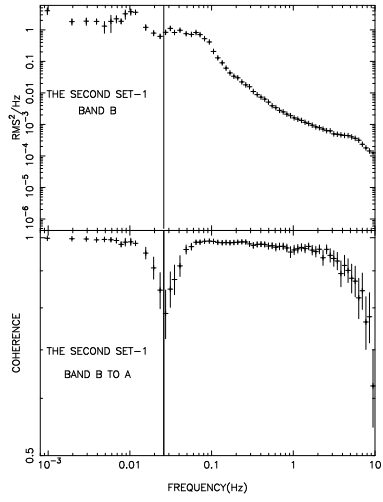}
\caption{The correlation between PDS and coherence function for the second data set.\label{fig3}}
\end{figure}

In GX339-4 (Nowak \etal  1999b), there is also a dip in the coherence function.
They suggested that there are multiple broad-band processes occurring in the source that are individually
coherent but they are incoherent relative to each other.
These dip frequencies are approximately the frequencies at which the different components overlap.
For the second set, we compare the coherence with the PDS: there are two features at frequencies below 1 Hz;
between these components, there is a dip in coherence (See figure ~\ref{fig3}); this is consistent to the above suggestion.

Nowak \etal  (1999a) proposed an accretion disk model for black hole binaries in which the corona is inside
the inner radius of the accretion disk, and
the soft photons from the accretion disk are up-scattered to hard photons in the corona. Therefore coherence loss
will occur at time scales shorter than the dynamical time scale of the corona. The break frequency in the coherence may then
be used to estimate the size of the corona, and thus the inner accretion disk radius.
For GRS1915+105, the characteristic frequency of about 0.02 Hz implies that the corona's size and the inner disk radius
is in the order of 10$^4$ GM/c$^2$ if the black hole is about 10$M_{\odot}$. This is clearly unphysical.

As discussed in Poutanen \etal  (1999), the X$/\gamma$-rays may be produced in compact magnetic flares at
radii $\leq$ 100 GM/$c^2$ from the central black hole. They predicted that the coherence will deviate from unity above a
characteristic frequency. This characteristic frequency is then associated with the longest time-scale of an
individual flare $\tau_{max}\simeq$1/2$\pi$$f_{br}$. If we take $f_{br}=0.02$ Hz,then $\tau$$_{max}$=8 s.
In fact the shape of the coherence curve is also remarkably similar to the prediction of their model.

\acknowledgments

We owe tremendously to the anonymous referee, who's deep insights, professionalism and invaluable suggestions have
improved the work significantly.
We thank interesting discussions with Prof. M. Wu and Dr. W. Yu. This study
is supported in part by the Special Funds for Major State Basic Research Projects and by the
National Natural Science Foundation of China.
SNZ also acknowledges supports
by NASA's Marshall Space Flight Center and through NASA's Long Term Space Astrophysics Program.

\clearpage

\begin{deluxetable}{cccc}
\tabletypesize{\scriptsize}
\tablecaption{The list of RXTE/PCA observations of GRS~1915+105 used for the analysis \label{Tb.1}}
\tablewidth{0pt}
\tablehead{
\colhead{Obs.ID} & \colhead{Data,UT}   & \colhead{Start,UT}   &
\colhead{Exp.,s}
}
\startdata
the first set &class $\phi$&& \\ \hline
10408-01-09-00 & 29/05/96 & 12:44 &5744 \\
10408-01-11-00 & 31/05/96 & 11:26 &10432 \\
10408-01-12-00 & 05/06/96 & 11:36 & 10600\\
10408-01-13-00 & 07/06/96 & 09:39 & 10832\\
10408-01-17-01 & 22/06/96 & 17:52 & 3392\\
10408-01-18-00 & 25/06/96 & 06:44 & 3680\\
10408-01-19-00 & 29/06/96 & 19:57 & 2160\\
10408-01-19-01 & 29/06/96 & 13:12 & 3344\\
10408-01-19-02 & 29/06/96 & 16:28 & 3184\\
10408-01-20-00 & 03/07/96 & 08:27 & 3328\\
10408-01-20-01 & 03/07/96 & 11:39 & 2936\\ \hline
the second set-1 &class $\mu$&&\\ \hline
10408-01-34-00 & 16/09/96 & 10:04 & 7920\\ \hline
the second set-2 &class $\mu$&&\\ \hline
10408-01-35-00 & 22/09/96 & 06:30 & 6624\\ \hline
the second set-3 &class $\mu$&&\\ \hline
10408-01-36-00 & 28/09/96 & 00:09 & 5600\\ \hline
the third set &class $\delta$&& \\ \hline
10408-01-14-00 & 12/06/96 & 00:06 &1312\\
10408-01-14-01 & 12/06/96 & 01:42 &1072\\
10408-01-14-02 & 12/06/96 & 03:18 &1072\\
10408-01-14-03 & 12/06/96 & 04:54 &1072\\
10408-01-14-04 & 12/06/96 & 06:30 &1408\\
\enddata
\end{deluxetable}

\begin{deluxetable}{cccc}
\tabletypesize{\scriptsize}
\tablecaption{Data modes, with their energy ranges, number of energy channels, and time resolution.\label{Tb.2}}
\tablewidth{0pt}
\tablehead{
\colhead{Data mode} & \colhead{Energy Range (keV)}   & \colhead{No.Chan.}   &
\colhead{Time Res (s)}
}
\startdata
B\_2ms\_4B\_0\_35\_H & 0-12.99 & 4 & $2^{-9}$ \\
E\_16us\_16B\_36\_1s & 12.99-100.0 & 16 & $2^{-16}$ \\
\enddata
\end{deluxetable}

\begin{deluxetable}{cccc}
\tabletypesize{\scriptsize}
\tablecaption{The energy bands used in this work (labeled A-E).\label{Tb.3}}
\tablewidth{0pt}
\tablehead{
\colhead{Energy Band} & \colhead{PHA Chan.}   & \colhead{Energy Range (keV)}   &
\colhead{Mean Energy (keV)}
}
\startdata
A & 0-13 & 0-5.12 & 3.94 \\
B & 14-18 & 5.12-6.89 & 5.96 \\
C & 19-25 & 6.89-9.39 & 7.79 \\
D & 26-35 & 9.39-12.99 & 10.95 \\
E & 36-79 & 12.99-29.26 & 16.97 \\
\enddata
\end{deluxetable}

\clearpage

\end{document}